\documentclass[english]{article}
\usepackage[T1]{fontenc}
\usepackage[latin9]{inputenc}
\usepackage{amsmath}
\usepackage{graphicx}
\usepackage{authblk}

\def\mbf{\mathbf}
\makeatletter

\@ifundefined{showcaptionsetup}{}{
 \PassOptionsToPackage{caption=false}{subfig}}
\makeatother

\usepackage{babel}
\begin{document}

\title{Topological transitions in evaporating thin films}

\author{Avraham Klein\thanks{avraham.klein@mail.huji.ac.il} }
\author{Oded Agam}
\affil{The Racah Institute of Physics, The Hebrew University, Jerusalem 91904, Israel}
%\ead{avraham.klein@mail.huji.ac.il}
\maketitle

\begin{abstract}
A thin water film evaporating from a cleaved mica substrate undergoes a first-order phase transition between two values of film thickness. During evaporation, the interface between the two phases develops a fingering instability similar to that observed in the Saffman-Taylor problem. The dynamics of the droplet interface is dictated by an infinite number of conserved quantities: all harmonic moments decay exponentially at the same rate. A typical scenario is the nucleation of a dry patch within the droplet domain. We construct solutions of this problem and analyze the toplogical transition occuring when the boundary of the dry patch meets the outer boundary. We show a duality between Laplacian growth and evaporation, and utilize it to explain the behaviour near the transition. We construct a family of problems for which evaporation and Laplacian growth are limiting cases and show that a necessary condition for a smooth topological transition, in this family, is that all boundaries share the same pressure.
\end{abstract}
\clearpage
\section{Introduction}
\label{sec:introduction}
  
The dynamics of evaporation and dewetting of thin liquid films are
important aspects of modern day technologies, such as in the
fabrication of electronic chips, microfluidic devices and
biosensors. These nonequilibrium processes depend on various details
such as the temperature distribution in the sample, the nature of the
interaction between the substrate and the fluid, and the substrate
roughness \cite{Bonn2009,Xue2011}. Depending on these details, a thin liquid film may, for
instance, break into isolated droplets, evaporate uniformly, or
exhibit a fingering instability similar to that observed in the
Saffman-Taylor problem \cite{Saffman1958} where an inviscid fluid
penetrates a viscous fluid. 

An experimental system showing this kind of behavior consists of a clean mica substrate
covered by a thin water film \cite{Elbaum1994,Lipson1996,Lipson2004}. Under proper conditions of temperature and pressure, two films of different
thickness (approximately 2 and 12 nm) can coexist. This behavior is a consequence of the competing van der
Waals and polar surface forces between the water and the substrate \cite{Samid-Merzel1998}. A first order phase
transition between the film heights is induced by changing the vapor pressure.
On the scale of nanometers, gravitational effects are negligible, and if the evaporation rate is sufficiently small one may neglect changes in the film thicknesses due to pressure gradients.
This implies that the dynamics of evaporation, in this system is, effectively, that of the interface between the phases.

An interesting phenomenon found in such a system is the nucleation of an `air' hole, such that an area of the thin film appears inside the thick film's area. This instability can then grow as the two boundaries of the thick phase move to meet one another. This can lead to breakup of the film into several components, or to the `expulsion' of the air-hole from within the droplet. In topological terms, the `droplet' (the thick phase) undergoes a topological transition from genus 0 to genus 1, which can then be followed by other transitions: either back to genus 0, or to a higher genus with some more complicated structure.

Recently, the evolution of evaporating thin films was related to Hele-Shaw type \cite{Richardson1972} dynamics \cite{Agam2009}, adding it to a long list of physical systems whose dynamics can be approximated in similar fashion, for example the so-called oil-well problem \cite{Kochina1996}, the thin-sheet stamping problem \cite{Entov1992-03-01}, and others. Such problems deal with the interface dynamics of two liquids with very different viscosities: A classic example is a droplet of a thin water film surrounded by air, which is a good description of the experimental setup described above. Many of these descriptions come under the general envelope of the idealized Laplacian growth problem. 

The connection to Hele-Shaw dynamics allows for the construction of an integrable model for the interface dynamics of the evaporating films. Then, one can devise explicit analytical solutions with which to conveniently explore the properties of such systems. See for example \cite{Mineev-Weinstein1994,Richardson1972,Entov1992-03-01,Crowdy2001-08-01} among many others. In this paper we shall construct such explicit solutions describing the topological transitions described previously for the evaporating films. We will discuss the effect of the transition on the boundary conditions necessary for a complete description of the problem. For example, adding an air-hole requires one to specify the pressure in the hole. This dramatically affects the behaviour of the interface as two boundaries approach one another and is closely related to previous work regarding idealized Laplacian growth problems (see e.g. \cite{Richardson1994}). Finally we shall characterize the transitions themselves and the dynamics close to such transitions. At this stage we will draw the connections with Laplacian growth problems. We will first demonstrate the duality of our system and Laplacian growth. Then we shall investigate how the stability of explicit solutions near the transitions depends on this duality. In the process we shall show how both Laplacian growth and thin film evaporation can be seen as special cases of a family of closely linked interface dynamics problems. This relationship becomes important when considering topological transitions.

The outline of this paper is as follows: In section \ref{sec:theoretical-model} we introduce the theoretical framework necessary to describe the phenomenon we are discussing. In section \ref{sec:meth-for-constr} we describe how to construct instructive solutions using methods borrowed from the idealized Laplacian growth problem. In section \ref{sec:exampl-expl-solut} we shall bring two concrete examples of such solutions. In section \ref{sec:char-trans} we shall examine the topological transitions in detail. Finally, in section \ref{sec:dual-lapl-growth}, we will draw connections between transitions in evaporation problems and in Laplacian growth problems. We shall use these connections to explain some of the features discussed in section \ref{sec:char-trans}.

\section{Theoretical model}
\label{sec:theoretical-model}

The first step to modeling our system is to divide it into two regions, according to the two phases described above. Since both phases have very small thickness, the dynamics are controlled by viscosity. By making a no-slip requirement along the substrate-liquid interface, and a no-shear requirement along the liquid-air interface, we can average over the vertical direction, arriving at the D'arcy law for the liquid velocity $\mbf{v}$:

\begin{equation}
{\mbf{v}} =-\frac{ h^2}{3\nu} \nabla P  \label{eq:darcy},
\end{equation}
where $h$ is the film height, $\nu$ is the viscosity, and $P=P(\mbf{r})$ is the average pressure at a point $\mbf{r}=(x,y)$. The liquid dynamics in the thin phase may be ignored, since as (\ref{eq:darcy}) shows the velocity there is approximately zero. We therefore call the thicker phase a `droplet' or water region and the thin phase a `dry' or air region. Henceforth we denote the droplet region by the symbol $D$. Next we assume that the film height in the droplet region is approximately fixed, allowing us to neglect variations in the two-dimensional liquid density $\rho$. Uniform evaporation then implies 
\begin{equation}
\rho \nabla \cdot {\mbf{v}} = - \kappa,  \label{eq:evaporation}
\end{equation}
where $\kappa$ is the rate of evaporation. Combining (\ref{eq:darcy}) and (\ref{eq:evaporation}) we see that the pressure obeys Poisson's equation in the droplet. 
For the time being, let us ignore the possible nucleation of an air-hole in the droplet. Then without loss of generality we may take the pressure in the dry domain to be constant and equal to zero. Thus:

\begin{equation}
  \label{eq:Poisson}
  \begin{array}{cc}
    \nabla^2 P = \beta & \mbox{inside~}D, \\
    P = 0 & \mbox{outside},
  \end{array}  
\end{equation}
where $\beta = 3\kappa\nu/\rho h^2$. From now on we shall assume units where it is dimensionless, along with both distance and time. A last simplification is to assume that surface tension can be neglected along the interface between regions so that the boundary condition on (\ref{eq:Poisson}) is:
\begin{equation}
  \label{eq:boundary}
  P(\mbf{r} \in \partial D) = 0.
\end{equation}
Here (and henceforth) $\partial D$ denotes the droplet boundary (for both simply- and doubly-connected droplets). The last assumption (\ref{eq:boundary}) has the effect of endowing our system with an infinite number of first integrals of motion \cite{Richardson1994,Entov1992-03-01}. Indeed, let $h$ be any harmonic function on $D$. Then we may define the following quantity:

\begin{equation}
  \label{eq:hMoment}
  t_h =\int_D h(\mbf{r})d^2r.
\end{equation}
We find the time dependence of $t_h$:

\begin{alignat}{3}
\label{eq:hMoment-time}
    \frac{d~t_h}{dt} & = & - &\oint_{\partial D} \frac{\partial P}{\partial n}h~dl \nonumber\\
  & = & - &\oint_{\partial D} \frac{\partial P}{\partial n}h~dl + \oint_{\partial D} P\frac{\partial h}{\partial n}~dl  \nonumber\\
  & = & &\int_D \left(- h\nabla^2P + P\nabla^2h \right)~d^2r \nonumber\\
  & = &  &\int_D -\beta h~d^2r = -\beta t_h
\end{alignat}
In this derivation we have used Green's theorem along with (\ref{eq:darcy}) and (\ref{eq:boundary}). $\partial/\partial n$ is the derivative perpendicular to the boundary. To utilize (\ref{eq:hMoment-time}), we define the so-called \emph{harmonic moments}:

\begin{align}
  \label{eq:harmonic-moments}
  t_k & =  \frac{1}{\pi}\int_D z^k dx~dy, \\
  t_k(t) & =  t_k(0)\exp(-\beta t).\nonumber
\end{align}
The time dependence of these moments can be used, in principal, to completely determine the interface dynamics. Let us clarify this through an example, by solving the problem of a circular droplet of radius $L$ with initial radius $L_0$. Solving (\ref{eq:Poisson})+(\ref{eq:boundary}) in polar coordinates $(r,\theta)$ we obtain $P(r,t) = \beta/4 [r^2-L(t)^2]$. In order to find $L(t)$ we note that $t_k = 0$ for any $k>0$ while $t_0 = L(t)^2 = L_0^2\exp(-\beta t)$. Thus $L(t) = L_0 \exp(-\beta t/2)$ solves the problem.
Now let us see what happens once we allow the creation of an air-hole in the droplet, which is the case this paper treats extensively. Let us use our previous example to illustrate. Assume that an air-hole of radius $l_0$ nucleates in the center of our circular droplet at $t=0$. The solution of (\ref{eq:Poisson}), (\ref{eq:boundary}) now changes to
\begin{equation}
\label{eq:pressure-annulus}
  P\left(r\right) = A+\frac{\beta}{4}r^{2}+\frac{B}{4}\log r
  \end{equation}
where $A  = -\beta l^{2}/4 -B\log l/4,  B = \beta\left(l^{2}-L^{2}\right)/\log (L/l)$. We see that adding the air-hole, i.e. changing the the droplet from being \emph{simply-connected} to being \emph{doubly-connected} has added a degree of freedom to the system, no longer determined by the harmonic moments \eqref{eq:harmonic-moments} alone. This is clearly a general aspect of such a change, not just an artifact of our example. We must constrain this freedom by adding an additional moment, which must express this topological change. This can be, for example, the logarithmic moment \cite{Richardson1994}
\begin{equation}
  \label{eq:log-moment}
  t_l = \frac{1}{\pi} \int_D \log\left|\mbf{r}\right| dx dy.
\end{equation}
$t_l$ decays exponentially like the other moments, since $log|\mbf{r}|$ is harmonic as long as $\mbf{0} \not\in D$ (as we shall assume henceforth for simplicity). In the case of our evaporating droplet, now in the shape of an annulus, this furnishes $t_l =L^{2}\log L-l^{2}\log l-\frac{L^{2}-l^{2}}{2}$. Adding this to $t_0=L^2-l^2$ gives two nonlinear equations for $l,L$, solving the problem. As expected, this solution depicts an annulus whose inner and outer boundaries approach each other. The solution exists for any finite $t$ and approaches a circle. One can find the circle's final radius by taking the limit $l\to L$, which yields $L(t\to \infty) = \exp[t_l(0)/t_0(0)]$. 
%Figure \ref{fig:annulus-evaporation} depicts an example of such an evaporating annulus. 
Of course, for such a simple geometry one can solve the problem by direct application of \eqref{eq:pressure-annulus} to D'Arcy's law \eqref{eq:darcy}, without recourse to moment equations \cite{Shelley1997}.

It should be noted, that when we claimed that adding an air-hole doesn't change the time dependence of the moments \eqref{eq:hMoment-time}, we implicitly used our assumption that the pressure is equal on both the inner and outer interfaces of the droplet, that is $\delta P = 0$ between boundaries. Although this certainly applies to the physical system we have in mind, one can readily think of cases where this is not so. For example, we can envision a system where \eqref{eq:Poisson} applies yet the pressure is zero only on one boundary. This could be a case where a droplet is compressed between two solid sheets at an exponential rate. An alternative is to envision that we take the limit $\beta \to 0$ but allow the pressure in the hole to be different from outside. This would be relevant for a Hele-Shaw cell, where we pierce the top sheet with a small hole and use that to enforce some pressure difference between boundaries. In this case, re-using the arguments in \eqref{eq:hMoment-time}, the time dependence of the moments becomes:
\begin{align}
  \label{eq:t_h_new}
  \frac{d~t_h}{dt} &= \int_D \left(- h\nabla^2P + P\nabla^2h \right)~d^2r -\frac{\delta P}{\pi} \oint_{\partial D_{out}} \frac{\partial h}{\partial n}dl\nonumber\\
  & = -\beta t_h - \frac{\delta P}{\pi} \int_{D_{out}} \nabla^2 h~d^2r.
\end{align}
Here $\partial D_{out}$ and $D_{out}$ are respectively the outer boundary of the droplet and the area it encloses. We have assumed without loss of generality that the pressure on the inner boundary is zero, while it is $\delta P$ on the outer boundary. Since $z^k$ are analytic in $D_{out}$, the $t_k$'s time dependence remains \eqref{eq:hMoment-time}, but now
\begin{equation}
  \label{eq:t_l_time_pressure}
  \frac{dt_l}{dt}=-\beta t_l -2\delta P.
\end{equation}
since $\nabla^2 \log(r) = 2\pi \delta^{(2)}(\mbf r)$.

Let us briefly summarize this section. We have simplified the problem of an evaporating droplet to solving Poisson's equation for the pressure, with $P = 0$ on the droplet boundary, for a moving boundary. This formulation leads to simple laws for the harmonic moments of the droplet, and these can be used to extract the time dependence of the boundary. A topological transition in the droplet, for example the appearance of an air-hole, adds a degree of freedom. This freedom has the physical interpretation of the ability to determine the pressure in the air-hole by various external means. It requires adding a constraint in the form of the logarithmic moment, whose time dependence encapsulates our new physical information on the system. Once this moment is known, we can again solve the given problems. 

For a more extended discussion of these points, and those in the following section, we refer the reader to the large body of work on Laplacian growth, for example \cite{Varchenko1992, Krichever2004} or the review \cite{Mineev-Weinstein2008}, as well as the works cited above. These also supply the connection between our problem and many other fields, which in the interest of brevity we do not address here.

\section{Method for constructing explicit solutions}
\label{sec:meth-for-constr}

\begin{figure}
  \centering
  \includegraphics[width=\linewidth,keepaspectratio]{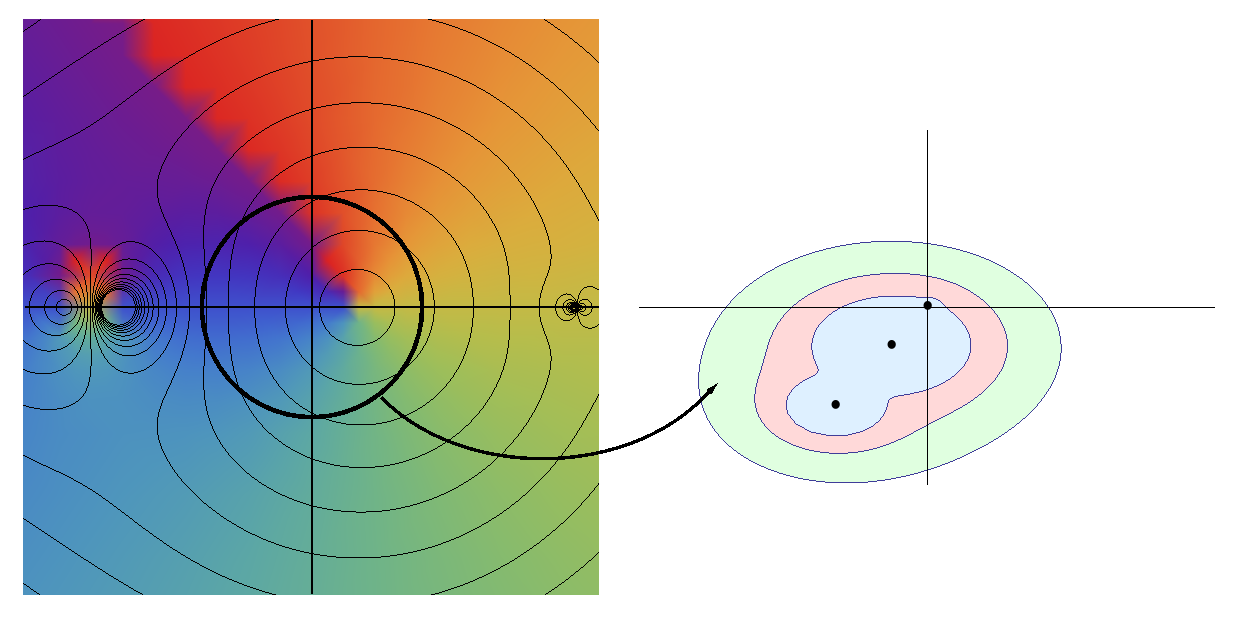}
  \caption{(color online) Mapping from the unit circle to a droplet. The left panel shows a mapping $z(w)$ that is conformal in the unit circle, having all its poles and branch points in the domain $|z|>1$. The contours represent lines of $|z(w)|=\mbox{Const}$ while the color shading denotes the phase $\mbox{Arg}~z(w)$. The right panel depicts how the unit circumference is mapped to the droplet' domain. Embedded in the droplet are the poles of the Schwarz function. Note: the left panel shows the mapping for the largest droplet contour.}
  \label{fig:genus-0-example}
\end{figure}
Unfortunately, finding the interface dynamics from the infinite number of moments $t_k$ (and $t_l$) is in general a formidable problem. It can be simplified greatly by assuming that the area of the droplet $D$ can be obtained as a conformal mapping from some simple region, with known singularities and branch cuts. For example we can choose a mapping so that our droplet domain $D$ (for now without a hole) is obtained as a mapping $z(w)$ from the unit circle $w<1$. In this way the moment constraints translate into equations for the parameters of $z(w)$. To see this we introduce the Schwarz function of $D$, defined as a function that is analytic in a strip about the boundary, and on the boundary obeys the relation $\bar{z} = S(z)$. For a mapping $z(w)$ as defined above, we have
\begin{equation}
  \label{eq:S_w_def}
  S(z)= \bar{z}\left[\frac{1}{w(z)}\right].
\end{equation}
Here $\bar{z}(w) = \overline{z({\bar{w})}}$ is the conjugate to $z(w)$, and the $1/w$ term appears since on boundary of the unit circle $\bar{w} = 1/w$, and we continue analytically from there. $w(z)$ is the inverse map of $z(w)$, which in general is multivalued, so that $S(z)$ is analytic on a (multi-sheeted) Riemann sphere . If $z(w)$ has poles of order $n_j$ at points $w_j$, by definition in the region $w > 1$, then $S(z)$ must have singularities at points $q_j\in D$ \emph{inside} the droplet, and obeying:
\begin{equation}
  \label{eq:s_params}
  q_j = z\left(\frac{1}{\overline{w_j}}\right), \qquad \mu_j = \lim_{\delta z \to 0} \delta z^{n_j} \bar{z}\left[\frac{1}{\overline{w_j}} + w'(q_j)\delta z + \ldots\right]
\end{equation}
This allows us to rewrite \eqref{eq:harmonic-moments} using Green's theorem as:
\begin{align}
  \label{eq:t_k_schwartz}
  t_k &= \frac{1}{2\pi i}\oint_{\partial D} z^k S(z) dz \nonumber\\
&= \sum_j \mu_j \binom{k}{n_j} q_j^{k+1-n_j}.
\end{align}
It is clear that in order to obey \eqref{eq:hMoment-time}, we must require that the poles $q_j$ be constant in time, while the weights $\mu_j$ all decay exponentially in time at the rate $\beta $:
\begin{align}
  \label{eq:pole-dependence}
  q_j = \mbox{const}, \qquad \mu_j(t) = \mu_j(0)e^{-\beta t}
\end{align}
Figure \ref{fig:genus-0-example} depicts an example of such a mapping.

\begin{figure}
  \centering
  \includegraphics[width=\linewidth,keepaspectratio]{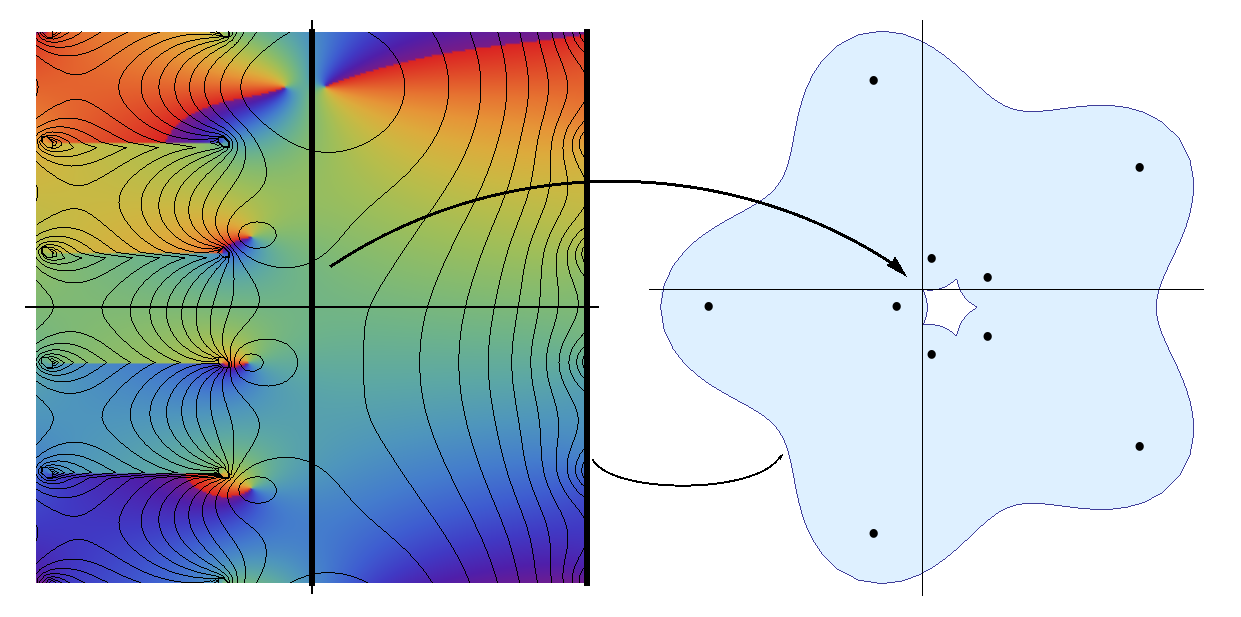}
  \caption{(color online) Mapping from a torus to a droplet with an air-hole. The left panel shows a mapping $z(u)$ that is analytic in the right half of the rectangle, having all its poles and branch points in the domain $u<0$. The contours represent lines of $|z(u)|=\mbox{Const}$ while the color shading denotes the phase $\mbox{Arg}~z(u)$. The mapping is elliptic, i.e. one should consider the top/bottom and left/right boundaries to be connected and forming a torus. The verticle lines $\Re u =0, \Re u = a_x$ are mapped to the droplet boundaries. Embedded in the droplet are the poles of the Schwarz function.}
  \label{fig:genus-1-example}
\end{figure}

\begin{figure}
  \centering
  \includegraphics[width=0.3\linewidth,keepaspectratio]{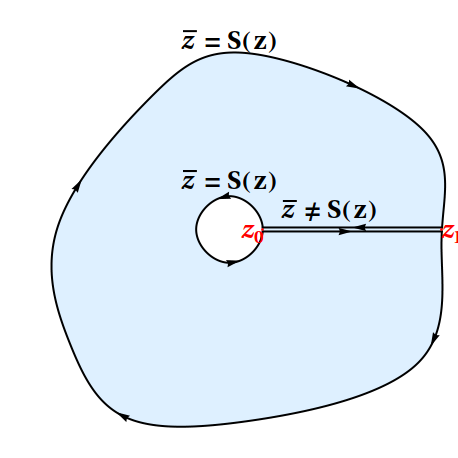}
  \caption{Integration path for the moments}
  \label{fig:integration-path}
\end{figure}

In order to incorporate an air-hole, we must find a mapping from a different domain. The most obvious would be to require that $D$ be obtained as a mapping $z(u)$ from some annulus $l<|u|<L$, such that each boundary of $D$ maps to a boundary of the annulus \cite{Richardson1996}. This adds a degree of freedom, namely the modulus of the annulus \cite{Nehari1952}, which is then constrained by $t_l$. In practice, we found it easier to work instead with a rectangle, so that $z(u)$ is an elliptic mapping from a rectangle of half-widths $a_x, a_y$. This means that $S(z) = \bar{z}(-\bar{u})$, so that $S(z)$ is analytic on a torus of modulus $\tau = a_y/a_x$. See figure \ref{fig:genus-1-example} for an example of such a mapping.
Now, to rewrite $t_k$ via Green's theorem we must introduce a cut in $D$, from some point on the inner boundary $z_0$ to some point on the outer boundary $z_1$ (see figure \ref{fig:integration-path}). It is easy to see that \eqref{eq:t_k_schwartz} is not changed by this addition. However, applying Green's theorem to $t_l$ requires care, as $D$ now winds around a singularity of $log(z)$. We write $t_l = \Re\frac{1}{\pi}\int_D \log z~d^2z$, and then using Green's theorem we have:
\begin{align}
  \label{eq:t_l_schwartz}
\frac{1}{\pi}\int_D \log z~d^2z & =  \frac{1}{2\pi i}\oint \log z \bar{z} dz \nonumber
\\ & = \frac{1}{2\pi i}\oint_{\partial D} \log z S(z) dz +\frac{1}{2\pi i}\left(\int_{z_1}^{z_0}+\int_{z_0}^{z_1}\right)\left[\bar{z} - S(z)\right] \log zdz
\end{align}
where the final term describes integrals along both branches of the cut (see figure \ref{fig:integration-path}). The first term on the right clearly depends only on $\mu_j,q_j$. To evaluate the second term we note that since $\log z = \log|z| +i \arg z $, the two integrals with the integrand $(\ldots)\log|z|$ exactly cancel, and with proper choice of the cut, one integral is on the branch $\arg z = 0$  and the other is on the branch $\arg z = 2\pi$. Thus:
\begin{align}
  \label{eq:t_l_schwartz_cont}
  t_l & = \Re \sum_j \mu_jR_j(q_j) -\Re \int_{z_0}^{z_1} [\bar{z} - S(z)] dz
\end{align}
Here $R_j$ are simply the appropriate residues of $\log z$. The first term  decays exponentially. The second term must decay at the same rate, and this provides the constraint on the torus modulus $\tau$ that is necessary to complete our system of equations. In fact, we may define the so-called `modified Schwarz potential' \cite{Shapiro1992,Crowdy2005}:
\begin{equation}
  \label{eq:Q_def}
  Q(z) = \frac{1}{4}\left[|z|^2 - \int^z S(z')dz' - \int^{\bar{z}} \bar{S}(z')dz'\right]
\end{equation}
named thus because it gives the contours $\bar{z} = S(z)$ as its saddle manifolds. Then, if we define a \emph{potential mismatch}\footnote{The use of the name `modified Schwarz potential' is perhaps slightly misleading. Usually this function is chosen so that both it and its derivative are zero on the boundaries (i.e. $\delta Q = 0$). However, the quantity we use captures the essential relation of the modified Schwarz potential to the Schwarz function, as defined in \cite{Shapiro1992}, so we use it here.}
\begin{align}
  \label{eq:dQ}
  \delta Q & = \Re\{ Q(z_1)-Q(z_0)\} = \frac{1}{2}\Re \int_{z_0}^{z_1}[\bar{z}-S(z)]dz \nonumber \\
& = \frac{1}{2}\left[\frac{1}{2}\left(|z_1|^2-|z_0|^2\right)- \Re \int_{z_0}^{z_1}S(z)dz\right]
\end{align}
we may use the following constraint instead of \eqref{eq:t_l_time_pressure}:
\begin{equation}
  \label{eq:dQ_t}
  \frac{d\delta Q}{dt} = -\beta \delta Q + \delta P
\end{equation}
since the second term on the right-hand side of \eqref{eq:t_l_schwartz_cont} is precisely $-2\delta Q$. Furthermore, let us recall that $S(z)$ is analytic on a Riemann surface of genus 1. It can be shown \cite{Krichever2005-10-01} that for a closed cycle $\Gamma$ on the surface, going from $z_0$ to $z_1$ in the droplet and then via an arbitrary route through the air region back from $z_1$ to $z_0$, one has:
\begin{align}
  \label{eq:dQ_riemann}
  \delta Q &= -\frac{1}{4}\oint S(z) dz. \nonumber \\
  &= -\frac{1}{4}\int_{u(\Gamma)}\bar{z}(-\bar{u})z'(u)du,
\end{align}
and we know that $u(\Gamma)$ begins on the line $\Re u =-a_x$ and goes to $\Re u = +a_x$. Since $S(z)$ is analytic on the surface, we may choose any loop homotopic to the one we chose, and we see that our choice of $z_0,z_1$ and $\Gamma$ was arbitrary. Note that the term \emph{potential mismatch} is apt in the sense that two boundaries cannot meet unless $\delta Q = 0$, since $\delta Q$ is defined in \eqref{eq:dQ} as an integral from one boundary to the other. Furthermore, if the pressure difference is zero then the mismatch decays as the boundaries approach. However, from \eqref{eq:dQ_t} we can immediately conclude that if there is a potential mismatch at the droplet's initial state, the two boundaries will not meet in any finite time.

\section{Examples of explicit solutions}
\label{sec:exampl-expl-solut}
\begin{figure}
  \centering
  \begin{minipage}{0.42\linewidth}
    \includegraphics[width=\linewidth,keepaspectratio]{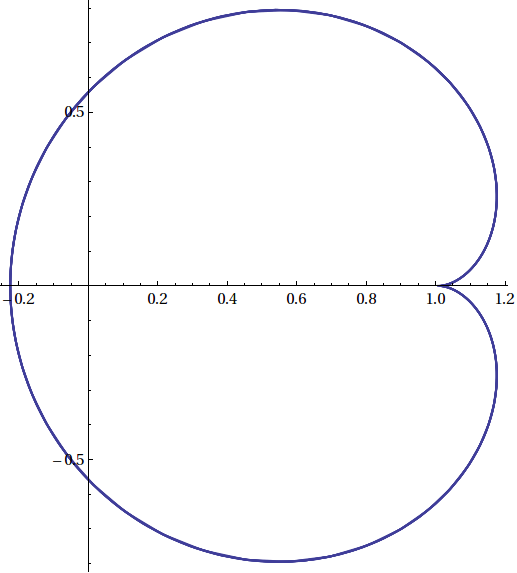}
  \end{minipage}
  \begin{minipage}{0.42\linewidth}
    \includegraphics[width=\linewidth,keepaspectratio]{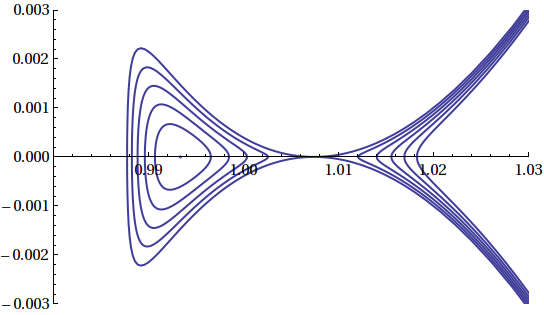}
  \end{minipage}

\caption{Nucleation of air-hole in a water droplet. The left panel shows the outline of the droplet defined by the mapping \eqref{eq:triple-pole-z}. Although at that magnification it can't be seen, there is a tiny air-hole on the $y=0$ axis near $x=1.01$. This can be seen in the magnified picture in the right panel. The contours in this figure denote equal-time steps in the evolution of the droplet domain. There are such contours in the left figure as well, but they are so closely spaced as to be indistinguishable. This shows how the short-scale dynamics of the air-hole evolution are much faster then that of the droplet as a whole.}
\label{fig:triple-single-figure}
\end{figure}

Using the method we described, one can construct solutions with features of interest. We shall present two examples of such solutions. The first can be seen as describing the nucleation of a small air-hole in the droplet, which then expands to meet the outer boundary, finally combining so that the final droplet is again simply connected. To this end we seek a Schwarz function whose behaviour in the droplet is of the form
\begin{align}
  \label{eq:triple-pole-schwartz}
  S(z) &= \sum_{j=0}^{2} \frac{\mu_j}{z-q_j} +\ldots \\
  q_j &\in D,\qquad \mu_i > 0. \nonumber
\end{align}
where $(...)$ denotes some analytic function. This form has been used previously in a number of interesting cases of doubly connected regions in various problems, e.g. \cite{RICHARDSON2000,Crowdy2002,CROWDY2003,Bettelheim2004}. The need for three poles can be explained qualitatively as follows: If $S(z)$ has one pole only, then the droplet it describes is highly symmetric, and there is no preferred spot for hole nucleation. The second pole therefore breaks this symmetry, and the third provides for the actual appearance of an air-hole. It is also possible to show formally \cite{Gustafsson1983-09-01, Bettelheim2004} that three poles is the minimum necessary for air-hole nucleation. To obtain such a mapping, we define an elliptic mapping with half periods $a_x, a_y$:
\begin{align}
  \label{eq:triple-pole-z}
  z(u) &= \sum_{j=0}^2 \alpha_j \zeta(u - u_j) + \gamma; \qquad -a_x < \Re u_j < 0,~-a_y < \Im u_j \leq a_y
\end{align}
where $\zeta(z)$ is the Weierstrass Zeta function, which is quasi-periodic. With properly chosen parameters, the mapping is conformal for $u$ in the right half of the fundamental region of the mapping, thus defining the droplet domain. To ensure periodicity we demand
\begin{align}
  \sum \alpha_j &= 0.\label{alpha-zeta-req},
\end{align}
and along with \eqref{eq:s_params}, \eqref{eq:t_l_schwartz_cont}, \eqref{eq:dQ_t} and \eqref{eq:dQ_riemann} this provides exactly 8 constraints for the seven parameters in the mapping plus the modulus $\tau = a_y/a_x$. These nonlinear equations can be solved numerically. An example of such a solution appears in Figure \ref{fig:triple-single-figure} \footnote{The parameters used here as initial conditions were, up to 5 digits of accuracy: $(u_0,u_1,u_2) = (-4.60281, -4.10607, -3.21041),~(\alpha_0,\alpha_1)=(0.99312,-1.28713), \gamma=0.99975,a_x=5.0,\tau = 1.2.$}.

\begin{figure}

  \centering
 \includegraphics[width=0.5\linewidth,keepaspectratio]{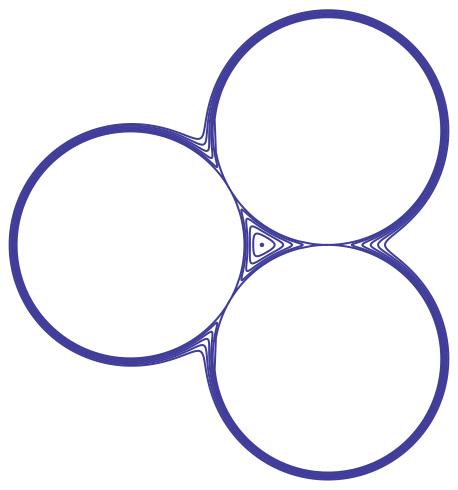}
  \caption{Breakup of a droplet as a result of air-hole nucleation. The figure denotes the evaporation of a droplet defined by a mapping of the form \eqref{eq:three-log-z}. The contours denote equal-time steps in the evolution of the droplet domain.}
\label{fig:log-figure}
\end{figure}

Our second example can be seen as describing how the nucleation of an air-hole ends up in breaking the droplet into separate fragments. We choose a mapping of the form \cite{Mineev-Weinstein1994,PonceDawson1998}:
\begin{align}
  \label{eq:three-log-z}
   \begin{split}
    z(u) & =\sum_{j=0}^2 \alpha_j \left[ \log\sigma(u - u_{j1}) - \log\sigma(u - u_{j2}) \right] + \gamma;\\
    &\qquad\qquad -a_x < \Re u_{jk} < 0,~-a_y < \Im u_{jk} \leq a_y
  \end{split}
  \\\Rightarrow S(z) &= \sum_{j=0}^2 \mu_j \left[ \log(z - q_{j1}) - \log(z - q_{j2}) \right]+\ldots
\end{align}
where $\sigma(z)$ is the Weierstrass Sigma function. It can be verified that the requirements \eqref{eq:pole-dependence} remain unchanged for a mapping of this form. The periodicity of $z(u)$ and the quasi-periodic behaviour of the Weierstrass Sigma function lead to the requirement,
\begin{align}
  \label{eq:log-periodicity-req}
  \sum_{j=0}^2 \alpha_j(u_{j1}-u_{j2}) = 0,
\end{align}
which along with the nine equations \eqref{eq:pole-dependence} plus \eqref{eq:dQ_t} fully determine the eleven parameters in \eqref{eq:three-log-z}. Figure \ref{fig:log-figure} denotes the evolution of such a droplet \footnote{The parameters used to start this solution were, to 6 digits of accuracy, $u_{jk}=-4.4+\delta_{k2}0.193+i\frac{14(j-1)}{3},\alpha_j = \exp\frac{2\pi i}{3}(j-1),\gamma=1.0,a_x = 5.0, \tau = 1.2.$}. For the examples shown in this figure, we have chosen a highly symmetric configuration, since otherwise it becomes very hard to numerically solve the large number of resultant nonlinear equations.

\section{Characterization of transitions}
\label{sec:char-trans}
\begin{figure}[t]
  \centering
  \includegraphics[width=\linewidth,keepaspectratio]{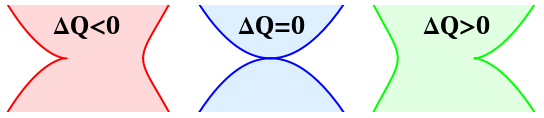}
  \caption{(color online) Cusp formation due to pertubation of initial conditions. The middle panel is a depiction of the final frame in figure \ref{fig:triple-single-figure}. The right and left panels depict the evolution of the droplet when $\delta Q \sim \pm 10^{-7}$. For comparison, all other parameters in this configuration take on values between an order of $10^{-3}$ to an order of $1$.}
  \label{fig:cusp-formation}
\end{figure}

In the previous section, we showed how explicit solutions exhibit features appearing in evaporating droplets. Both these examples showed how an evaporating droplet eventually undergoes a topological transition from genus 1 to genus 0. This contrasts sharply with the example of an evaporating annulus from section \ref{sec:theoretical-model}, where the droplet remains in the form of an annulus at any finite time - a clearly unphysical solution resulting from symmetry. One mathematical reason for this can be found immediately. For both examples in section \ref{sec:exampl-expl-solut}, the initial conditions were chosen so that the potential mismatch $\delta Q = 0$ initially (and thus forever). For the annulus, we can find the mismatch by noting that the Schwarz function for an evaporating annulus has, near the boundaries, the form $R^2/z$, where $R=l,L$ on the inner and outer boundaries respectively. Using this, one can easily identify that for the annulus $t_l = - \delta Q$. This value is clearly non-zero as can be seen from the calculation in section \ref{sec:theoretical-model}. However, this difference is not the whole story. Figure \ref{fig:cusp-formation} depicts the result of pertubating the initial conditions of the mapping shown in figure \ref{fig:triple-single-figure}, so that $\delta Q \not = 0$ but is still very small compared to all other parameters in the problem. As can be seen in the figure, the result is a formation of a cusp on one boundary (leading to a breakdown of the solution) with the other boundary remaining regular. We note that a similar behaviour to the one shown in figure \ref{fig:cusp-formation} was found in a different context in \cite{Richardson1994}. In addition, for this specific example, the velocity at the regular boundary goes to zero as the cusp forms. This behaviour is very similar to what one would expect were there actually a pressure difference on the boundaries.% Another aspect of this solution is that the velocity of the interface at the regular boundary goes to zero. We believe that an explanation for such behaviour is necessary.

A simple explanation for this behaviour can be found by noting that as can be seen from \eqref{eq:Q_def}, the function $\beta Q(z)$ itself solves Poisson's equation in the droplet domain. It cannot however be identified with the pressure, since it has singular points associated with the singularities of $S(z)$. Nevertheless, near the transition area, far from the singular points, $Q$ itself behaves very nearly like a pressure field itself, with the important addition that $\nabla Q = 0$ on the boundary, i.e. the boundaries are saddle manifolds of $Q$. When $\delta Q = 0$, then there is no problem for these saddle manifolds to remain stable until the actual transition when the boundaries meet. However, if $\delta Q \not = 0$, then one of two options must occur: either the dynamics slow down to ensure that the saddle points remain stable, or if the interface velocity is too high then at some point one of the saddle manifolds, associated with the higher value of $Q$, must become unstable, leading to a breakdown of the solution. Thus $Q$ acts as a quasi-pressure field itself, and the difference in this quasi-pressure causes the breakdown of the solution. 

We shall now investigate this pressure analogy further.

\section{Duality of Laplacian growth and evaporation}
\label{sec:dual-lapl-growth}

In this section we shall give the modified Schwarz potential $Q$ an interpretation as a real pressure field. We shall do this by utilizing a duality between thin film evaporation and the idealized Hele-Shaw cell, or Laplacian growth, problem. By duality we mean that for a given evaporation problem with some known solution, we can construct an analogous Laplacian growth problem leading to a different pressure field but with the same solution for the interface dynamics. This relation has also been studied in the context of the `inverse balayage' problem in potential theory and quadrature domains (see e.g. \cite{Gustafsson1997} and \cite{Gustafsson2002}). The pressure field in this dual problem will naturally contain $Q$ as one of its elements. We will then generalize this duality to a family of analogous problems, all featuring the exact same interface dynamics. This generalization will demonstrate how a potential mismatch, i.e. $\delta Q \not=0$, in an evaporation problem, translates to a pressure difference in the analogous problems, without affecting the interface dynamics, and explains the behaviour shown in figure \ref{fig:cusp-formation}.

Let us briefly cover the definition of the idealized Hele-Shaw cell evolution problem. Suppose that instead of evaporation, the droplet dynamics are governed by the existence of pumps at discrete points $q_j$, such that in the vicinity of these points the pressure diverges as $P \sim \log|z-q_j|$. In this case, the pressure obeys the following relation:
\begin{align}
  \label{eq:pressure-hele-shaw}
  \nabla^2\tilde{P} = \sum_j R_j(t) \delta^{(2)}(z - q_j)
\end{align}
where $R_j(t)$ denote the rate of pumping at the points $q_j$. Positive $R_j$ are equivalent to liquid extraction, and negative $R_j$ to injection. In the case of a simply connected droplet, it is possible to reduce our evaporation problem to Laplacian growth, simply by placing sinks with rates $R_j(t) = \beta \pi \mu_j(t)$ at the poles of the Schwarz function. This can be verified by directly calculating the harmonic moments by methods analogous to those used in \eqref{eq:hMoment-time} and \eqref{eq:t_k_schwartz}. However, in the case of a multi-connected droplet, in order to obtain the same evolution for the potential mismatch, and thus the same dynamics, one must apply a pressure difference of $-\beta \delta Q(t)$ on the boundary. This can be seen from \eqref{eq:dQ_t}, by taking the limit $\beta \to 0$. Thus we see that the dynamics of an evaporating droplet with \emph{potential} mismatch $\delta Q$ are the same as a Hele-Shaw droplet with a \emph{pressure} mismatch $-\beta \delta Q$. In fact, by using the correspondence of the singularities of $S$ and $Q$, as is evident from \eqref{eq:Q_def}, we can readily identify that the solution to \eqref{eq:pressure-hele-shaw} with the relevant boundary is simply
\begin{align}
  \label{eq:tilde-p-1}
  \tilde{P} = P-\beta Q.
\end{align}
With this, the fact that the dynamics are the same for this droplet is trivially obtained from $Q$'s saddle-manifold behaviour: the interface velocity is precisely $\mbf{v} \propto -\nabla \tilde{P} = -\nabla P$. For $\delta Q \not= 0$ the pressure difference is non-zero and so we know that the boundaries cannot meet. In this case it is less clear how the pressure difference causes the cusp formation, except that one expects that the area of higher pressure will actually be more regular (as opposed to the evaporation problem) as is indeed the case. We still don't have a clear picture of why the regular boundary should slow down.

We can take this point further by trying to construct more problems featuring the same dynamics. To do so, we can envision a situation where we have both evaporation and pumping as evolution mechanisms. Let us concentrate on the case where we are moving liquid via pumps, at an exponentially decreasing rate, while evaporating at a different rate, so that the pressure obeys the condition

\begin{align}
  \label{eq:tilde-p-2}
  \nabla^2 \tilde P = \beta + \alpha \left[1 - \sum_j \pi \mu_j(t)\right] \delta^{(2)}(z- q_j)
\end{align}
so that the droplet is evaporating at a rate $\beta + \alpha$, but the $\alpha$ evaporation is being exactly canceled by injection (or extraction). Clearly $\alpha = 0$ reduces to our original problem, and $\alpha = -\beta$ reduces to Laplacian growth. Now, for a harmonic function $h$ we have, by adapting \eqref{eq:t_h_new} to our new system
\begin{equation}
  \label{eq:t_h_new_2}
  \begin{split}
    \frac{d~t_h}{dt} &= \int_D \left(- h\nabla^2\tilde P + \tilde P\nabla^2h \right)~d^2r -\frac{\delta \tilde P}{\pi} \oint_{\partial D_1} \frac{\partial h}{\partial n}dl\\
  &= \int_D -h\left\{\beta  + \alpha \left[1 - \sum_j \pi \mu_j(t)\right] \delta^{(2)}(z- q_j)\right\} d^2r + \\
  &\quad+ \int_D \tilde P\nabla^2h~ d^2r -\frac{\delta \tilde P}{\pi} \int_{D_1}\nabla^2h~ d^2r~.
\end{split}
\end{equation}
For the moments $t_k$, i.e. $h = z^k/\pi$, the last two terms on the RHS fall and the first simply evaluates to $-\beta t_k$, from \eqref{eq:t_k_schwartz}. However, for $t_l$, that is $h = \pi^{-1}\Re \log z$, we have:
\begin{equation}
  \label{eq:t-l-new}
  \frac{d~t_l}{dt} = -(\beta + \alpha) t_l + \alpha \sum_j \mu_j(t) \log q_j -2\delta\tilde P.
\end{equation}
Using \eqref{eq:t_l_schwartz_cont} we can rewrite this as:
\begin{equation}
  \label{eq:t-l-new-final}
  \frac{d~t_l}{dt} = -(\beta + \alpha) t_l + \alpha (t_l - 2\delta Q) -2\delta\tilde P,
\end{equation}
and so in order to get the same dynamics as the evaporating droplet we need:
\begin{equation}
  \label{eq:dq-new}
  \alpha\delta Q(t) = \delta \tilde P(t)
\end{equation}
So, for any $\alpha \not=0$ we see that applying a very specific pressure difference will yield the same dynamics. Note, that in this case the pressure has to be determined from $Q$ and not vice-versa. (This is in close analogy to the case discussed in \cite{Krichever2005-10-01}.) In addition, we see that any $\alpha >0$ is equivalent to having a \emph{positive pressure difference} which is exactly the behaviour that explains the shapes in figure \ref{fig:cusp-formation}. For any $\alpha < 0$ we have the equivalent of negative pressure difference. In that sense, the question of whether the pumps are extracting or injecting liquids appears to determine the mechanism of breakdown. Finally, it is again clear that the solution to \eqref{eq:tilde-p-2} is precisely $\tilde P = P + \alpha Q$. In this sense $Q$ represents the pressure field resulting from evaporation along with a compensating injection. Our original analogy indeed manifests itself in a concrete manner.

We should emphasize a couple of points. First of all, we have not shown that a pressure mismatch $\delta Q \not=0$ necessarily leads to one of the boundaries coming to a halt. The solution to a given problem could well break down long before one boundary comes to a halt. In other words, a breakdown will occur when a saddle contour of $Q$ becomes unstable, but that doesn't mean that the interface velocity will have reached zero by then. We have also not shown that $\delta Q = 0$ must lead to a meeting of boundaries even if the only important dynamics are occuring near the transition point. Perhaps cusps appear before the meeting. We have only shown that the mismatch leads to a bound on the time for which a solution exists, and provided a mechanism for the breakdown. The actual exact relation between a solution's breakdown and the evolution of $Q$ remains a matter for further study.

\subsection*{Acknowledgements}
We would like to thank E. Bettelheim and D. Khavinson for very helpful discussions. We would also like to thank the anonymous referees for their helpful and detailed comments. This research was supported by the Israel Science Foundation (ISF) grant No. 9/09.

\bibliographystyle{unsrt}
\bibliography{evaporation}
\end{document}